\documentclass[12pt,final]{article}
\usepackage{graphicx}
\usepackage{epsfig}
\usepackage{dcolumn}
\usepackage{bm}

\usepackage{amssymb,amsmath}
\usepackage{gensymb}
\usepackage{multirow}
\usepackage{booktabs}
\makeatletter
\def\@biblabel#1{(#1)}
\makeatother
\usepackage{setspace}

\usepackage{booktabs,rotating}
\usepackage{fancyhdr}
\usepackage{booktabs,caption}
\usepackage[flushleft]{threeparttable}
\usepackage{enumerate,subfigure,tabularx,longtable}
\usepackage{longtable}
\usepackage[utf8x]{inputenc}
\UseRawInputEncoding
\usepackage{gensymb}
\usepackage {threeparttable} 
\newcommand{\et}{\textit{et al.}}

\newcommand{\comment}[1]{}

\def\gsim {\mbox{\hbox{ \lower-.6ex\hbox{$>$}
\kern-1.12em \lower.5ex\hbox{$\sim$}\kern+.35em}}}
\def\lsim {\mbox{\hbox{ \lower-.6ex\hbox{$<$}
\kern-1.12em \lower.5ex\hbox{$\sim$}\kern+.35em}}}
\makeatletter
\let\@fnsymbol\@arabic
\makeatother

\begin{document}


\title{\vspace{-3.0cm} 
	Resolving Discrepancies in Disjoining Pressure Predictions for Liquid Nanofilms from Molecular Simulations}

	\author{Yafan Yang\thanks{To whom correspondence should be addressed, email: yafan.yang@cumt.edu.cn} $^{,\dag,\ddag}$, 
		Zufeng Zuo\thanks{To whom correspondence should be addressed, email: zufeng\_zuo@163.com} $^{,\dag}$, \\ 
		Jingyu Wan$^{\dag}$, Shuyu Sun$^{\S}$, 
		and Denvid Lau\thanks{To whom correspondence should be addressed, email: denvid.lau@cityu.edu.hk} $^{,\ddag}$\\
		\\[-15pt] 
		\normalsize  $^{\dag}$State Key Laboratory of Intelligent Construction and Healthy Operation \\
		\normalsize  and Maintenance of Deep Underground Engineering, \\
		\normalsize  China University of Mining and Technology, 
		\normalsize  Xuzhou 221116, Jiangsu, China. \\
		\normalsize $^\ddag$Department of Architecture and Civil Engineering, \\
		\normalsize  City University of Hong Kong, 
		\normalsize  Hong Kong 999077, China. \\
		\normalsize $^\S$School of Mathematical Sciences, Tongji University, Shanghai 200092, China.\\
	}

\date{\today}
\maketitle
\newpage
\begin{abstract}
Literature values of disjoining pressure in liquid nanofilms from different molecular simulation methods show significant discrepancies. We demonstrate that these arise from neglecting long-range dispersion interactions and inconsistent definitions of film thickness in the original Peng method. A key insight is that long-range dispersion affects surface tension in a thickness-dependent manner, increasing it at large thickness but suppressing its enhancement at small thickness due to disjoining-pressure-induced normal compression and lateral expansion. This leads to crossover behavior in the surface tension of water nanofilms. Since disjoining pressure is obtained from the derivative of surface tension with respect to thickness, this nontrivial dependence strongly impacts its accuracy. With proper treatment of dispersion interactions and a consistent thickness definition, the revised Peng method agrees with the Bhatt method and yields more accurate Hamaker constants.
\end{abstract}

\clearpage

\section{Introduction}

Liquid films with thicknesses of a few nanometers are commonly found in systems such as foams and emulsions, where the films are bounded by vapor on both sides. These thin films play an important role in many applications, including flotation and separation processes, coating technologies, and georesources exploitation.\cite{albijanic2010review,li2023shale} Under equilibrium conditions, the stability of such systems is closely related to the magnitude of the disjoining pressure within the film.\cite{chatzigiannakis2021thin}

\begin{figure}[tb]
\begin{centering}
	\includegraphics[width=0.9\textwidth]{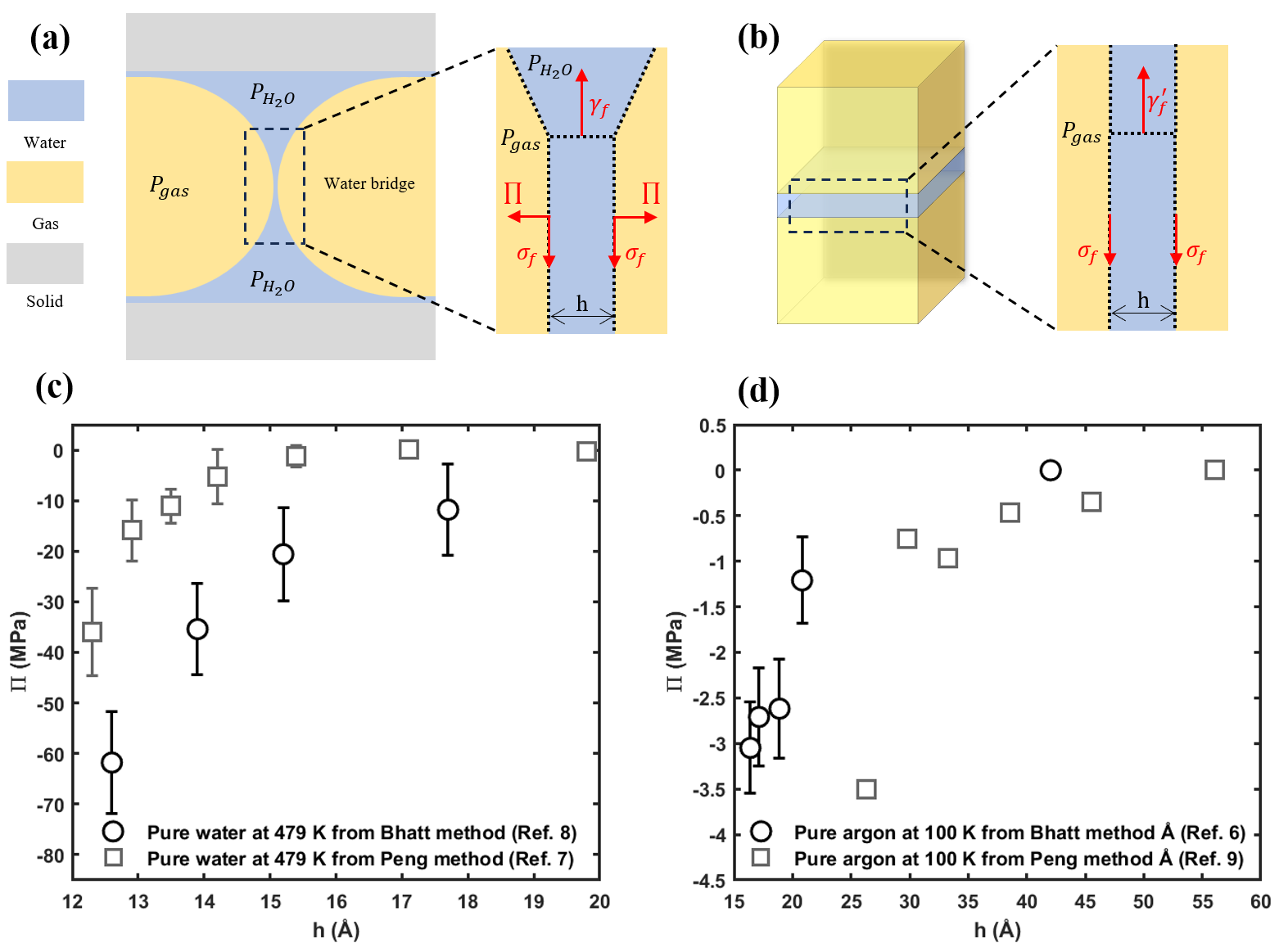}
	\caption{(a) Schematic representation of a thin water film confined in a pore. The film is in equilibrium with the bulk water through a Plateau-border meniscus and is surrounded by gas phases on both sides. The corresponding force balance acting on the film is also illustrated. 
		(b) System setup for estimating the surface tension $\sigma_f$ and the corresponding force balance of the thin film in the absence of a meniscus.
		Large discrepancies observed in disjoining pressures calculated by Bhatt method\cite{bhatt2003molecular,bhatt2002molecular} and Peng method\cite{peng2015methodology,peng2016modelling} for (c) pure water and (d) pure argon systems.
	}
	\label{fig:z1}
\end{centering}
\end{figure}

Fig. \ref{fig:z1}a illustrates a schematic representation of a gas-saturated water thin film confined within a pore. The film is in equilibrium with the bulk liquid through a Plateau-border meniscus and is bounded by gas phases on both sides. The disjoining pressure $\Pi$ is defined by the following equation:\cite{ivanov2023thin,derjaguin1936anomalien}
\begin{equation}
\label{eq:0}
\Pi=P_{gas}-P_{H_2O},
\end{equation}
where $P_{gas}$ is the bulk pressure of the gas phase, which equals to the normal pressure within the film, and $P_{H_2O}$ represents the pressure of the homogeneous liquid  forming the constant-curvature region of the meniscus at the same chemical potential as the film.

Such film systems can be unstable due to long-wavelength fluctuations and are therefore difficult to investigate experimentally.\cite{bhatt2002molecular} Molecular simulations have provided invaluable insight into the behavior of disjoining pressure in liquid nanofilms, where the finite size of the simulation box suppresses these fluctuations and stabilizes the films. There are mainly two approaches developed by Bhatt \et\cite{bhatt2002molecular} and Peng \et\cite{peng2015methodology} for calculating the disjoining pressure isotherms. 

Bhatt \et\cite{bhatt2002molecular} first proposed a method based on an integrated Gibbs-Duhem equation. In this approach, the chemical potential is calculated using Widom's insertion method, and the disjoining pressure can also be obtained by extrapolation from a chemical potential-pressure phase diagram using an analytic equation of state. 
This approach requires the chemical potential to be evaluated in the low-density vapor region, where Widom's insertion method is more reliable than in the liquid phase. Consequently, its application to low-temperature systems is limited because the coexistence vapor density becomes negligibly small.

To address this limitation, Peng \et\cite{peng2015methodology} developed a new method based on film thermodynamics. In this approach, a relation was used in which the disjoining pressure is proportional to the derivative of the surface tension with respect to the film thickness. This method is advantageous because the surface tension can be readily obtained from molecular simulations at various temperatures.

It is important to note that in both methods the same system setup is employed, consisting of a slab of liquid film sandwiched between vapor phases in the absence of a meniscus, as shown in Fig. \ref{fig:z1}b. Both methods have been applied to determine the disjoining pressure isotherms of pure water\cite{bhatt2003molecular,peng2015methodology} and pure argon systems.\cite{bhatt2002molecular,peng2016modelling} Comparisons of the results are shown in Fig. \ref{fig:z1}c and d. 

Surprisingly, the two approaches show poor agreement, despite the use of the same system setup and the simplicity of the systems studied. The discrepancy is significantly larger than the uncertainties associated with either method. For the water system, the disjoining pressure predicted by the Peng method is nearly twice that obtained using the Bhatt method at the smallest film thickness. As the film thickness increases, the results from the Peng method rapidly approach zero at around 15.4~\AA, whereas the Bhatt method still predicts a value of about -10 MPa at a thickness of approximately 32.7~\AA. An even larger discrepancy is observed for the argon system. Only qualitative agreement can be obtained between the two approaches.

In this work, we aim to resolve this large discrepancies by refining the Peng method. We demonstrate that two adjustments to the original Peng method are necessary to reconcile these differences. First, the long-range interactions of the Lennard-Jones (LJ) potential (i.e., the dispersion term) must be incorporated into the calculation of surface tension. Second, a compatible definition of the film thickness should be adopted. A detailed discussion of these two points is provided below through a brief description of the Peng method.

\section{Method}
Based on the film thermodynamics, the following equation describing the relations between film tension $\gamma_f$, surface tension $\sigma_f$, disjoining pressure, and film thickness $h$ can be derived:\cite{ivanov2023thin,toshev1975thermodynamics}
\begin{equation}
\label{eq:1}
\gamma_f=2\sigma_f+ \Pi h.
\end{equation}

Eq.~\ref{eq:1} indicates that the film tension, measured at the hypothetical sharp plane separating the thin film and the meniscus, consists of two contributions, as shown in Fig.~\ref{fig:z1}a, the surface tensions of the two film interfaces and an additional term arising from interactions across the film thickness, represented by the disjoining pressure.

In addition, another relation can be obtained:\cite{ivanov2023thin}
\begin{equation}
\label{eq:2}
\gamma_f=2\sigma_{\infty}+ \int_{h}^{\infty} \Pi dh +  \Pi h,
\end{equation}
where $\sigma_{\infty}$ is the surface tension of an film with infinite thickness ($i.e., h=\infty$). 

Equating Eq. \ref{eq:1} and Eq. \ref{eq:2}, the following relation is obtained:\cite{peng2015methodology}
\begin{equation}
\label{eq:3}
\int_{h}^{\infty} \Pi dh = 2\sigma_f - 2\sigma_{\infty}.
\end{equation}

The above equation shows that the difference in the work required to stretch a thin film and a thick film per unit area is equal to the reversible work per unit area required to bring the two film surfaces into contact.\cite{eriksson1982disjoining}

Finally, taking the derivative of Eq. \ref{eq:3} with respect to $h$ yields:\cite{peng2015methodology}
\begin{equation}
\label{eq:4}
\Pi = -2\cfrac{\partial\sigma_f}{\partial h}.
\end{equation}

This equation serves as the key relation for calculating the disjoining pressure in molecular simulations. The same equation was previously derived by Ivanov and Toshev.\cite{ivanov1975thermodynamics2}

The surface tension of the film can be estimated using molecular simulations with the configuration illustrated in Fig. \ref{fig:z1}b. In this setup, the molecular system consists of a liquid film slab surrounded by the gas phase. Although the Plateau-border meniscus is absent in this configuration, the simulated film remains in the same equilibrium state because the chemical potential is maintained equal to that of the bulk gas phase. As the tension plane does not sense the presence of the meniscus in such a system, Eq. \ref{eq:1} reduces to:
\begin{equation}
\label{eq:5}
\gamma_f'=2\sigma_f,
\end{equation}
where $\gamma_f'$ is the film tension measured in center regions of the film.

This enables the direct evaluation of the surface tension through Bakker's equation:\cite{bakker1928kapillaritat,green1960molecular}
\begin{equation}
\label{eq:6}
\sigma_f = \cfrac{\gamma_f'}{2}=\cfrac{1}{2}\int_{-\infty}^{+\infty} \Big[P_{zz}-\cfrac{1}{2}(P_{xx}+P_{yy})\Big] dz,
\end{equation}
where $P_{xx}$, $P_{yy}$, and $P_{zz}$ are the three diagonal components of the pressure tensor along the $x$, $y$, and $z$ direction with $z$ direction normal to the interface. 

Note that in the Peng method, the surface tension of pure water\cite{peng2015methodology} and argon\cite{peng2016modelling} was calculated without accounting for the long-range contribution of the LJ interaction, a simplification that is commonly adopted in molecular simulations.\cite{berendsen1987missing,bhatt2002molecular} In contrast, although surface tension is not explicitly calculated when determining the disjoining pressure in the Bhatt method, Bhatt \et\cite{bhatt2003molecular} included the long-range LJ contribution in their simlations of the water system. For the argon system, however, a large cutoff distance was used and interactions beyond the cutoff were neglected.\cite{bhatt2002molecular}

The primary motivation of the present study is to elucidate the effects of long-range dispersion interactions on the interfacial properties of liquid nanofilms.
It is well known that neglecting long-range LJ interactions leads to an underestimation of surface tension.\cite{alejandre1995molecular,werth2013influence,stephan2020influence} 
The magnitude of this underestimation has been examined over a range of temperatures by Alejandre \et\cite{alejandre1995molecular} They reported that neglecting long-range LJ interactions leads to a systematic underestimation of the surface tension of SPC/E water by approximately 9\% over the temperature range 328-573 K.
If the underestimation is also systematic with respect to film thickness, its influence on the disjoining pressure obtained from the Peng method may be limited, because the Peng method depends on the variation of surface tension with film thickness rather than its absolute value.
Nevertheless, whether this approximation contributes to the large discrepancy between the two methods remains unclear, as the influence of long-range LJ interactions on the surface tension of thin liquid films has not yet been directly investigated. 

Another motivation is to identify an appropriate definition of film thickness for estimating the disjoining pressure in the Peng method.
Different definitions of film thickness have been reported in the literature.\cite{ivanov1975thermodynamics2,bhatt2002molecular,bhatt2003molecular,peng2015methodology}
The definition of film thickness can significantly influence the disjoining pressure calculated using the Peng method, since it appears as the differentiation variable (see Eq. \ref{eq:4}).
In the original work proposing the Peng method, the film thickness was defined based on the system size obtained from bulk liquid water simulations performed in the isothermal-isobaric ensemble.\cite{peng2015methodology,peng2015surface} The film configuration was then constructed by introducing two vacuum regions on either side of the bulk liquid. However, this approach raises concerns. First, the imposed pressure is not explicitly known, and choosing an inconsistent value can significantly affect the calculated film thickness, particularly for highly compressible liquids. Second, the presence of vacuum regions may induce evaporation of liquid molecules into the vapor phase. If the resulting vapor density is non-negligible, a considerable number of molecules may leave the liquid region, effectively reducing the liquid content of the film and leading to an overestimation of the film thickness. Therefore, a consistent and physically well-defined definition of film thickness is essential.

In fact, the thermodynamically consistent definition of the film thickness, consistent with Eqs.~\ref{eq:1}-\ref{eq:4}, was proposed earlier by Ivanov and Toshev:\cite{ivanov1975thermodynamics2}
\begin{equation}
\label{eq:7}
h=\cfrac{N/A-\rho_vL_z}{\rho_l-\rho_v},
\end{equation}
where $N$ is the number of molecules in the system, $L_z$ is the box length in $z$-direction, $\rho_l$ and $\rho_v$ are the bulk densities of liquid and vapor phases, respectively. In this definition, two dividing surfaces are introduced to represent the two film-vapor interfaces, and the film thickness is defined as the separation between these surfaces. Conceptually, the dividing surfaces partition the system into hypothetical fluid regions with uniform density corresponding to that of the bulk liquid in the meniscus. Because the meniscus is not explicitly simulated, the bulk liquid density $\rho_l$ is approximated by the density measured in the central region of the thickest simulated film. The assumption of a constant liquid density is justified by the relatively low compressibility of the liquid under the selected conditions.\cite{linstrom2001nist,bhatt2002molecular,bhatt2003molecular} Note that this definition was also adopted in Bhatt method.\cite{bhatt2002molecular,bhatt2003molecular}

All molecular dynamics simulations in this work were conducted using the LAMMPS code.\cite{thompson2022lammps} A fully flexible version of the SPC/E water model was employed because it provides a better description of surface tension.\cite{lopez2008effect,berendsen1987missing}
The argon molecular model was taken from Sherwood and Prausnitz.\cite{sherwood1964intermolecular} 
To investigate the effect of long-range interactions of the LJ potential on the calculated disjoining pressure, two cases were considered. In the first case, the long-range LJ interactions was ignored and the cutoff distances for the argon and water systems were set to 22.776~\AA\ and 16.500~\AA, respectively. In the second case, long-range LJ interactions were treated using the particle-particle-particle-mesh (PPPM) method with a relative error of 10$^{-4}$. For charged systems, long-range Coulombic interactions were evaluated using the PPPM method with a relative error of 10$^{-4}$. 

To facilitate comparison with literature data,\cite{bhatt2002molecular,bhatt2003molecular,peng2015methodology} simulation box sizes similar to those used in previous studies were adopted. The cross-section sizes for argon and water systems are 49.056~\AA$\times$49.056~\AA\  and 36.000~\AA$\times$36.000~\AA, respectively. The box length in the \( z \)-direction is set to be three times that in the other directions. Periodic boundary conditions were applied in all directions. The pure component systems were simulated in the NVT ensemble with temperature being controlled using Nos\'e-Hoover thermostat. The coupled Newton’s equations of motion were integrated using the velocity Verlet algorithm. For the argon system, the equilibration and production periods were 7.5 ns and 22.5 ns, respectively, with a time step of 5 fs. For the water systems, the equilibration and production periods were 3 ns and 6 ns, respectively, with a time step of 1 fs. The production run was divided into three blocks for uncertainty estimation.

\section{Results and Discussion}

\begin{figure}[tb]
\begin{centering}
	\includegraphics[width=0.5\textwidth]{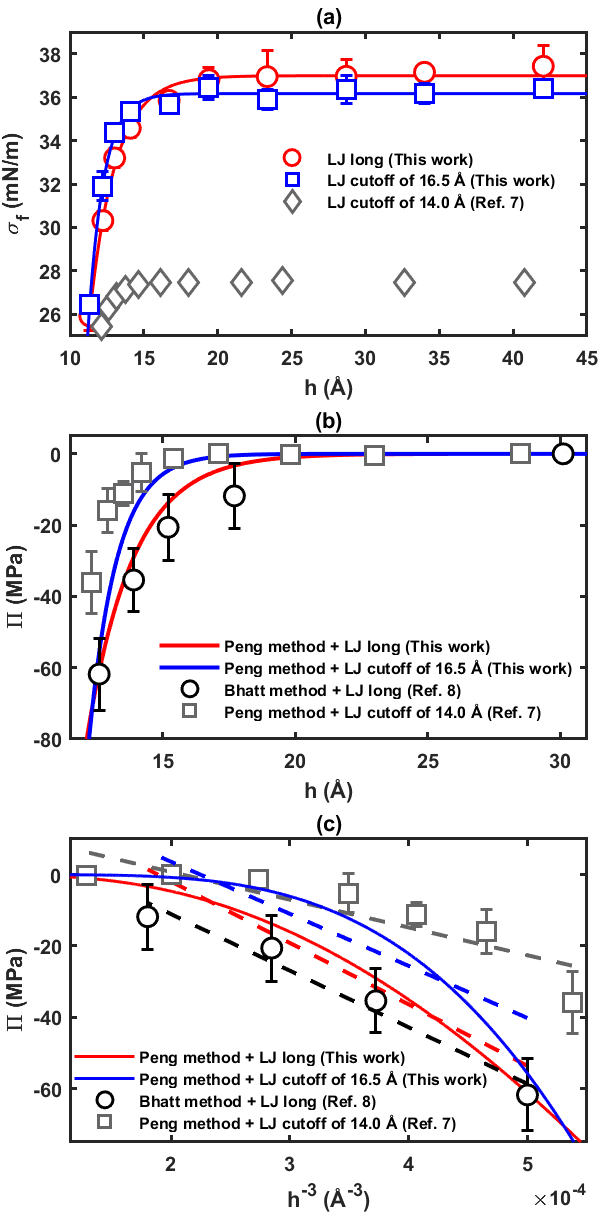}
	\caption{(a) Surface tension of the water nanofilm as a function of film thickness at 479 K. Solid lines represent the fitted curves. 
		(b) Corresponding disjoining pressure as a function of film thickness.
		(c) Corresponding disjoining pressure as a function of the inverse cubic film thickness. 
		Dashed lines represent linear fits.
		Our results are compared with those reported by Peng \textit{et al.}\cite{peng2015methodology} and Bhatt \textit{et al.}\cite{bhatt2003molecular}.
	}
	\label{fig:z2}
\end{centering}
\end{figure}

Fig.~\ref{fig:z2}a shows the calculated surface tension of water as a function of film thickness at 479~K using different treatments of the long-range dispersion interaction, with corresponding data provided in Table~1. The experimental surface tension at this temperature is 36.3~mN/m. 
At the largest film thickness, our calculated surface tensions are $36.4 \pm 0.34$~mN/m and $37.4 \pm 0.94$~mN/m when the LJ dispersion interaction is treated using a cutoff of 16.5~\AA\ and the PPPM method, respectively. These values are in good agreement with the experimental result. 
For comparison, Bhatt \et\cite{bhatt2003molecular} reported a surface tension of $33.5 \pm 3.1$~mN/m when long-range dispersion interactions were included, while Peng \et\cite{peng2015methodology} obtained a significantly lower value of $27.47 \pm 0.10$~mN/m using a cutoff of 14.0~\AA. 
The improved agreement of our results with the experimental value can be attributed to the inclusion of bond and angle flexibility in the water model.\cite{lopez2008effect} In addition, these results further confirm that truncating the dispersion interaction with a finite cutoff tends to underestimate the surface tension.\cite{alejandre1995molecular}

The surface tension increases with film thickness. However, distinct plateau values are observed for cases with and without long-range interactions. For the cutoff cases, the variation in surface tension is primarily confined to film thicknesses below 15 \AA. When long-range interactions are included, the plateau shifts to a higher thickness of around 20 \AA. Nevertheless, the overall difference between the two cases in our simulations remains moderate, with a maximum deviation of approximately 5\%.

An important finding of this work is that the influence of long-range dispersion interactions on water surface tensions depends strongly on the film thickness. Opposite effects are observed on the two sides of a crossover point at a film thickness of approximately 15.5~\AA.
At large film thickness, the inclusion of long-range dispersion interactions increases the calculated surface tension, which is consistent with previous literature reports. This behavior is expected because truncating dispersion interactions weakens the attractive intermolecular forces that contribute to the in-plane contraction of the interface, thereby leading to an underestimation of the surface tension.

In contrast, at small film thickness the inclusion of long-range dispersion interactions leads to a decrease in the surface tension. This unexpected behavior can be understood from the role of disjoining pressure. As the film thickness decreases, the disjoining pressure of water increases in magnitude. For pure water films, the disjoining pressure is negative and therefore acts as a driving force for film thinning.\cite{bhatt2003molecular,peng2015methodology} Including long-range dispersion interactions also strengthens this thinning force in the direction normal to the film. The resulting compression of the water film in the normal direction generates an expansion tendency in the lateral direction. 
Although the in-plane contraction force from surface tension also increases when long-range interactions are included, this increase is smaller than the enhanced lateral expansion induced by the disjoining pressure. Consequently, the net effect is a reduction in the surface tension at small film thickness.

The disjoining pressure-film thickness relationship can be described by an exponential function:\cite{li2024hygroscopic,peng2017surface}
\begin{equation}
\label{eq:8}
\sigma_f=a \cdot e^{b\cdot h}+c,
\end{equation}
where $a$, $b$, and $c$ are the fitting parameters in units of $mN/m$, $\AA^{-1}$, and $mN/m$. Substituting Eq. \ref{eq:8} into Eq. \ref{eq:4} yields the following expression for the disjoining pressure:
\begin{equation}
\label{eq:9}
\Pi=-20a\cdot b\cdot e^{b\cdot h}. 
\end{equation}
Here, $\Pi$ is in unit of MPa. A factor of 10 is needed for unit conversion. 

The fitted surface tension values are presented as solid lines in Fig. \ref{fig:z2}a. All the fitting parameters are provided in Table 2. Overall, the fitted curves show good agreement with the simulation data, with the Pearson correlation coefficients of 0.997 and 0.974 for cases with and without long-range dispersion interactions, respectively.

Fig.~\ref{fig:z2}b shows a comparison of the disjoining pressure.
Our results obtained with long-range dispersion interactions are in good agreement with those from the Bhatt method\cite{bhatt2003molecular}, with deviations remaining within the uncertainty of the data. In contrast, the results based on cutoff treatment are more consistent with the data reported by Peng \et\cite{peng2015methodology}. These findings confirm that long-range dispersion interactions play a crucial role in reconciling the discrepancies between the two methods.

The classical Hamaker theory gives:\cite{israelachvili2011intermolecular}
\begin{equation}
\label{eq:10}
\Pi = -\frac{A_H}{6\pi h^3},
\end{equation}
where \( A_H \) is the Hamaker constant. To examine this relationship, \( \Pi \) is plotted as a function of \( 1/h^3 \) in Fig.~\ref{fig:z2}c, and linear fits are performed based on Eq.~\ref{eq:10}. 
Overall, the disjoining pressures obtained from our simulations with long-range dispersion, as well as those reported by Bhatt \et\cite{bhatt2003molecular}, exhibit better agreement with the linear scaling predicted by the Hamaker theory compared to results based on cutoff treatments. The fitted Hamaker constants are \( 3.2 \times 10^{-18} \)~J and \( 2.8 \times 10^{-18} \)~J for our simulations with and without long-range dispersion, respectively. Corresponding values reported by Bhatt \et\cite{bhatt2003molecular} and Peng \et\cite{peng2015methodology} are \( 3.0 \times 10^{-18} \)~J and \( 1.5 \times 10^{-18} \)~J, respectively.
The relatively close agreement between our results with long-range dispersion and those of Bhatt \et\cite{bhatt2003molecular} suggests that properly accounting for long-range interactions is essential for recovering the physically meaningful Hamaker constants. In contrast, the smaller value obtained using cutoff treatments indicates that truncation of dispersion interactions leads to an underestimation of the effective van der Waals attraction. This trend is consistent with the observed deviation from the expected \( h^{-3} \) scaling in such cases.
Moreover, the Lifshitz theory predicts that \( A_H \approx 6.2 \times 10^{-20}\)J for water at 479~K.\cite{dagastine2000dielectric,bhatt2003molecular} This value is approximately 50 times smaller than those obtained from simulations, and this discrepancy has been attributed to the neglect of fluid structure, which becomes particularly important at nanometer-scale film thicknesses.\cite{bhatt2003molecular}

\begin{figure}[tb]
\begin{centering}
	\includegraphics[width=0.95\textwidth]{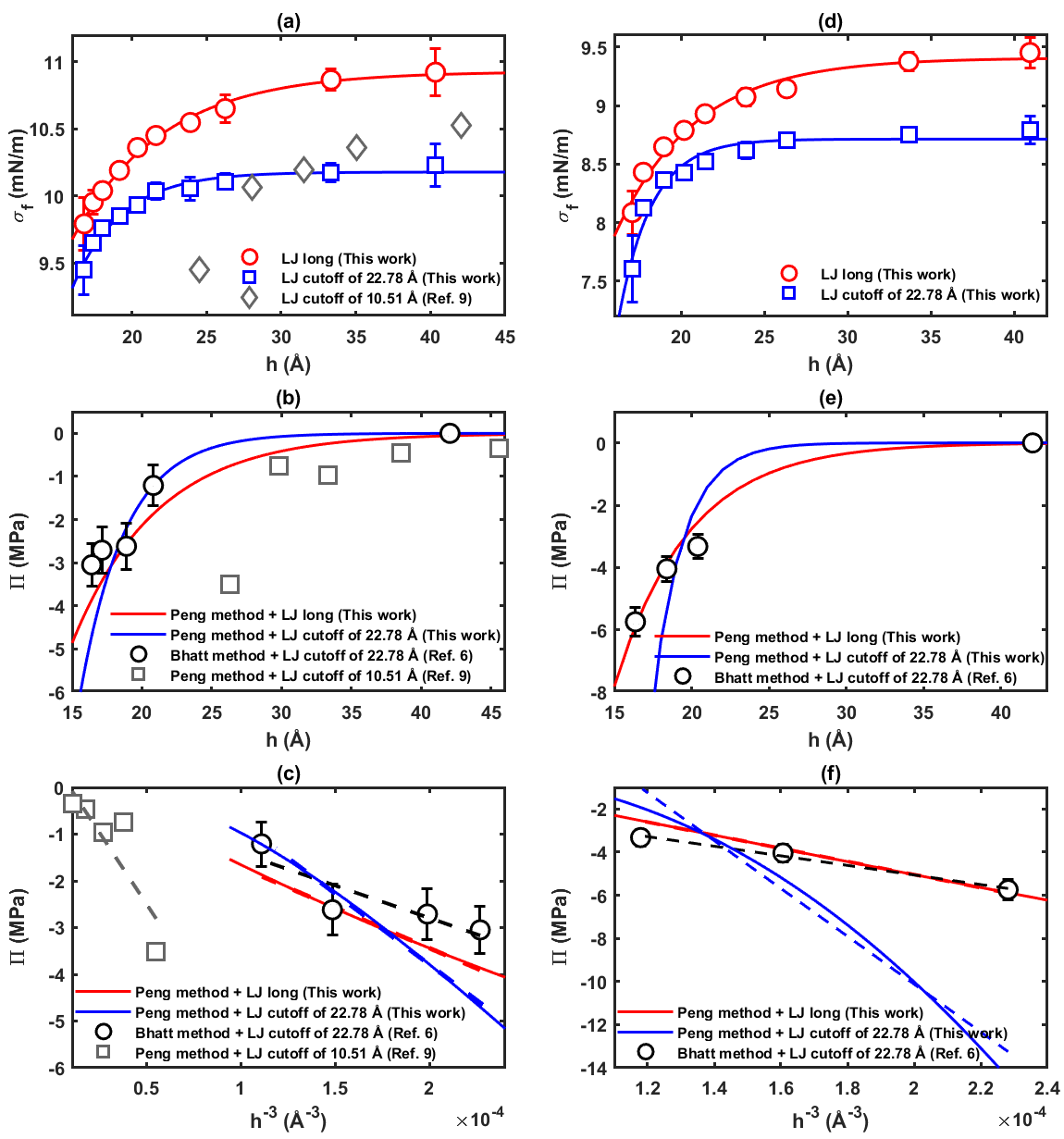}
	\caption{Top panels: Surface tension of the argon nanofilm as a function of film thickness. Solid lines in top panels represent the fitted curves.
		Middle panels: Corresponding disjoining pressure as a function of film thickness.
		Bottom panels: Corresponding disjoining pressure as a function of the inverse cubic of the film thickness. Dashed lines represent linear fits.
		Left and right panels correspond to temperatures of 100.05 K and 105.93 K, respectively. 
		Our results are compared with those reported by Peng \textit{et al.}\cite{peng2016modelling} and Bhatt \textit{et al.}\cite{bhatt2002molecular}.}
	\label{fig:z3}
\end{centering}
\end{figure}

We further investigated liquid argon nanofilms at 100.05 K and 105.93 K. Figs. \ref{fig:z3}a and \ref{fig:z3}d present the calculated surface tensions obtained with different treatments of long-range dispersion interactions. It is evident that neglecting long-range dispersion leads to a systematic underestimation of the surface tension across the entire range of film thicknesses. Notably, this discrepancy is significantly larger than that observed in the water system. 
This stronger sensitivity can be attributed to the nature of intermolecular interactions in the two systems. In SPC/E water, long-range Coulombic interactions are explicitly accounted for using the PPPM method, while the LJ contribution plays a relatively secondary role. In contrast, argon is described solely by the LJ potential, in which dispersion interactions dominate the intermolecular attraction. As a result, the treatment of long-range LJ interactions has a much more pronounced effect on the calculated surface tension in argon than in water.

Unlike the water system, the crossover behavior associated with opposite effects of long-range dispersion at small and large film thicknesses is not observed for argon. This difference can be attributed to the relatively small magnitude of the disjoining pressure in argon films (see Figs. \ref{fig:z3}b and \ref{fig:z3}e), which leads to a weaker normal stress across the film. As a result, the coupling between normal compression and lateral expansion, which is responsible for the reduction of surface tension in thin water films, is much less pronounced in argon.
Nevertheless, a slight reduction in the discrepancy between the two treatments is observed at smaller film thicknesses. This trend suggests that thickness-dependent effects begin to emerge, although they remain insufficient to offset the dominant contribution of long-range dispersion interactions to the surface tension.

We also compared our simulated surface tensions with those reported by Peng et al.\cite{peng2016modelling}. Our results reach a plateau at smaller film thicknesses. This difference is mainly attributable to the definition of film thickness. As discussed above, the film thickness in the Peng method is affected by the imposed pressure as well as by evaporation from the liquid film.

Figs.~\ref{fig:z3}b and \ref{fig:z3}e present the calculated disjoining pressures at two temperatures (the equations of the curves are provided in Table S3). Overall, our results including long-range LJ dispersion show much better agreement with those obtained using the Bhatt method than our results based on a cutoff treatment.\cite{bhatt2002molecular}
Although Bhatt\et\cite{bhatt2002molecular} employed a cutoff scheme, the observed agreement is likely due to the relative insensitivity of the chemical potential to long-range dispersion interactions, which is used to estimate the disjoining pressure.
In contrast, the results from the Peng method\cite{peng2016modelling} deviate significantly from the other approaches, primarily due to differences in the definition of film thickness. Notably, the agreement between our long-range dispersion results and the Bhatt method improves at higher temperatures. This trend is likely related to the increased difficulty of accurately calculating the chemical potential at lower temperatures.\cite{bhatt2002molecular} 

Generally, the disjoining pressures obtained for the argon systems exhibit a clear linear dependence on \( h^{-3} \), consistent with the prediction of the Hamaker theory, as shown in Figs.~\ref{fig:z3}c and ~\ref{fig:z3}f. The extracted Hamaker constants at 100.05 K are \( 3.3 \times 10^{-19} \) J and \( 5.7 \times 10^{-19} \) J for our simulations with and without long-range dispersion, respectively, compared to \( 2.6 \times 10^{-19} \) J reported by Bhatt \et\cite{bhatt2002molecular} and \( 1.2 \times 10^{-18} \) J from the Peng method.\cite{peng2016modelling} At 105.93 K, the corresponding values are \( 5.7 \times 10^{-19} \) J and \( 2.1 \times 10^{-18} \) J, while Bhatt \et\cite{bhatt2002molecular} reported \( 4.2 \times 10^{-19} \) J. 
The relatively good agreement between our results with long-range dispersion and those of Bhatt \et\cite{bhatt2002molecular} suggests that an accurate treatment of dispersion interactions is important for obtaining physically meaningful Hamaker constants. In contrast, the significantly larger values obtained without long-range corrections and from the Peng method indicate that methodological differences, particularly in the treatment of dispersion interactions and the definition of film thickness, can lead to substantial overestimation of the effective van der Waals attraction. 
Furthermore, the Lifshitz theory predicts a Hamaker constant of \( 1.6 \times 10^{-20} \) J at 105.93 K,\cite{bhatt2002molecular} which is one to two orders of magnitude smaller than the simulation results. This discrepancy is consistent with previous observations and can be attributed to the neglect of molecular-scale structural effects in continuum theory, which becomes increasingly important for thin films at nanometer length scales.

\section{Conclusion}
In this work, we clarify the origin of the large discrepancies in disjoining pressures of liquid nanofilms reported in previous molecular simulation studies. The results show that these discrepancies arise from neglecting long-range dispersion interactions and from inconsistencies in the definition of film thickness.
A central finding is that long-range dispersion interactions exert a thickness-dependent effect on surface tension. At large film thickness, their inclusion strengthens intermolecular attraction and increases the surface tension, consistent with conventional understanding. However, at small film thickness, they enhance the magnitude of the disjoining pressure, leading to stronger normal compression and a corresponding lateral expansion force, which reduces the net increase in surface tension. This competition can even give rise to a crossover behavior in the water system.
This thickness dependence is critical because the Peng method relates disjoining pressure to the derivative of surface tension with respect to film thickness. Consequently, even moderate variations in the thickness dependence of surface tension can lead to significant differences in the predicted disjoining pressure. When long-range dispersion interactions are properly included, the calculated disjoining pressures for both water and argon systems agree well with those obtained from the Bhatt method, confirming the importance of accurately accounting for dispersion interactions.

We further show that the definition of film thickness plays a crucial role in the Peng method. The original definition can introduce systematic errors due to unknown pressure conditions and evaporation effects, whereas a thermodynamically consistent definition based on bulk densities provides a physically meaningful measure of film thickness and yields consistent results across methods.
Moreover, disjoining pressures obtained with proper treatment of long-range interactions more closely follow the expected $h^{-3}$ scaling and yield Hamaker constants in good agreement with those from the Bhatt method. In contrast, cutoff-based treatments lead to noticeable deviations and overestimation of effective van der Waals interactions.
Overall, this study demonstrates that the interfacial thermodynamics of liquid nanofilms is governed by a subtle coupling between long-range intermolecular interactions, film thickness, and mechanical balance within the film. These findings provide a clear physical interpretation of previously reported discrepancies and establish a consistent framework for predicting disjoining pressure, with direct implications for nanoscale interfacial phenomena in foams, emulsions, and confined fluids.


\bigskip
\bigskip
{\bf{ACKNOWLEDGMENTS\\[1ex]}}
The research is supported by the Fundamental Research Funds for the Central Universities (2025QN1175).

\bibliography{DisP}


\clearpage

\begin{table}[!ht]
\centering\caption{Surface tension data of liquid nanofilms from molecular simulations with long-range dispersion treated by PPPM method ($\sigma_f^{PPPM}$) and cutoff technique ($\sigma_f^{cutoff}$).}
\begin{tabular}{c|c|c|c|c|c|c}
	\hline
	System & Temp. (K) & $h$ (\AA) & $\sigma_f^{PPPM}$  (mN/m) & Error & $\sigma_f^{cutoff}$  (mN/m) & Error \\ \hline
	Water & 470.00  & 11.34  & 25.92  & 0.65  & 26.45  & 0.28  \\ 
	Water & 470.00  & 12.23  & 30.31  & 0.44  & 31.91  & 0.68  \\ 
	Water & 470.00  & 13.02  & 33.20  & 0.45  & 34.37  & 0.24  \\ 
	Water & 470.00  & 14.08  & 34.57  & 0.44  & 35.33  & 0.07  \\ 
	Water & 470.00  & 16.71  & 35.85  & 0.37  & 35.67  & 0.27  \\ 
	Water & 470.00  & 19.41  & 36.81  & 0.54  & 36.44  & 0.54  \\ 
	Water & 470.00  & 23.35  & 36.95  & 1.21  & 35.89  & 0.47  \\ 
	Water & 470.00  & 28.70  & 36.96  & 0.78  & 36.34  & 0.63  \\ 
	Water & 470.00  & 34.00  & 37.14  & 0.22  & 36.17  & 0.45  \\ 
	Water & 470.00  & 42.04  & 37.42  & 0.94  & 36.41  & 0.34  \\ \hline
	Argon & 100.05  & 16.81  & 9.79  & 0.20  & 9.45  & 0.18  \\ 
	Argon & 100.05  & 17.41  & 9.95  & 0.09  & 9.65  & 0.03  \\ 
	Argon & 100.05  & 18.04  & 10.04  & 0.03  & 9.76  & 0.04  \\ 
	Argon & 100.05  & 19.17  & 10.19  & 0.02  & 9.85  & 0.05  \\ 
	Argon & 100.05  & 20.38  & 10.36  & 0.06  & 9.93  & 0.04  \\ 
	Argon & 100.05  & 21.61  & 10.45  & 0.02  & 10.03  & 0.06  \\ 
	Argon & 100.05  & 23.92  & 10.55  & 0.05  & 10.05  & 0.09  \\ 
	Argon & 100.05  & 26.25  & 10.65  & 0.10  & 10.11  & 0.06  \\ 
	Argon & 100.05  & 33.35  & 10.87  & 0.08  & 10.17  & 0.07  \\ 
	Argon & 100.05  & 40.33  & 10.92  & 0.17  & 10.23  & 0.16  \\ 
	Argon & 105.93  & 17.08  & 8.09  & 0.18  & 7.60  & 0.29  \\ 
	Argon & 105.93  & 17.74  & 8.43  & 0.03  & 8.13  & 0.03  \\ 
	Argon & 105.93  & 18.97  & 8.65  & 0.02  & 8.36  & 0.03  \\ 
	Argon & 105.93  & 20.19  & 8.79  & 0.04  & 8.42  & 0.04  \\ 
	Argon & 105.93  & 21.45  & 8.93  & 0.05  & 8.52  & 0.04  \\ 
	Argon & 105.93  & 23.90  & 9.07  & 0.07  & 8.62  & 0.07  \\ 
	Argon & 105.93  & 26.35  & 9.14  & 0.06  & 8.70  & 0.05  \\ 
	Argon & 105.93  & 33.68  & 9.38  & 0.08  & 8.75  & 0.04  \\ 
	Argon & 105.93  & 40.96  & 9.45  & 0.13  & 8.79  & 0.12 \\ \hline
\end{tabular}
\end{table}

\clearpage
\begin{sidewaystable}
\centering\caption{Fitted results for liquid nanofilms.}
\begin{threeparttable}
	\begin{tabular}{c|c|c|c|c|c|c}
		\hline
		System & Temp. (K) & Method & $\sigma_f$ (mN/m) with $h$ in \AA & $\Pi$ (MPa) with $h$ in \AA& $R$ for $\sigma_f$ & $H_A$ (J) \\ \hline
		Water & 479.00 & PPPM & $-7263.3128e^{-0.5726h}+36.9859$ & $-83179.7049e^{-0.5726h}$ & 0.997 & $3.2 \times 10^{-18}$  \\ 	\hline
		Water & 479.00 & Cutoff & $-455653.2328e^{-0.9485h}+36.1638$ & $-8643477.9302e^{-0.9485h}$ & 0.974 & $2.8 \times 10^{-18}$  \\ 		\hline
		Argon & 100.05 & PPPM & $-17.4375e^{-0.1646h}+10.9282$ & $-57.3870e^{-0.1646h}$ & 0.997 & $3.3 \times 10^{-19}$  \\ 	\hline
		Argon & 100.05 & Cutoff & $-115.8424e^{-0.3059h}+10.1779$ & $-708.6332e^{-0.3059h}$ & 0.988 & $5.7 \times 10^{-19}$  \\ \hline			
		Argon & 105.93 & PPPM & $-42.8370e^{-0.2086h}+9.4060$ & $-178.7188e^{-0.2086h}$ & 0.989 & $5.7 \times 10^{-19}$  \\ 	\hline
		Argon & 105.93 & Cutoff & $-5170.8272e^{-0.5000h}8.7128$ & $-51703.4099e^{-0.5000h}$ & 0.975 & $2.1 \times 10^{-18}$  \\ 				\hline			  
	\end{tabular}
\end{threeparttable}
\end{sidewaystable}

\clearpage

\clearpage

\clearpage
\begin{figure}[tb]
\begin{centering}
	\centerline{\includegraphics[width=1.1\textwidth]{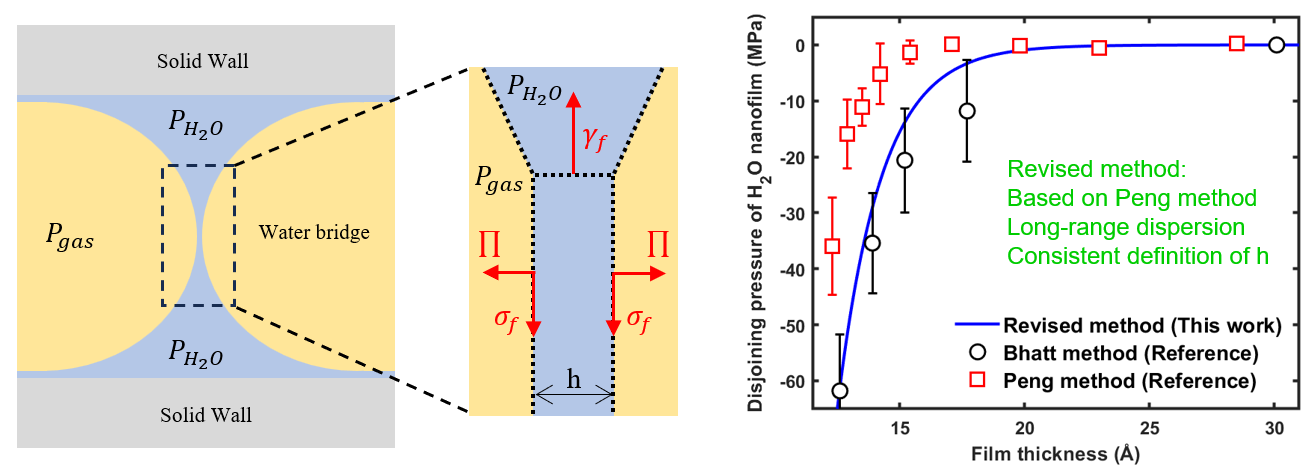}}
	\vskip0.5cm
\end{centering}
\vskip3cm
{\large \bf TOC Graphic\\[2ex]}
\end{figure}

\end{document}